\documentclass[english]{article}
\usepackage[T1]{fontenc}
\usepackage[latin9]{inputenc}
\usepackage{geometry}
\usepackage{color}
\usepackage{amsmath}
\usepackage{amsthm}
\usepackage{wasysym}

\makeatletter
\theoremstyle{plain}

\theoremstyle{remark}

\usepackage{babel}

\makeatother

\usepackage{babel}
\providecommand{\remarkname}{Remark}
\providecommand{\theoremname}{Theorem}

\begin{document}
\title{\textbf{Second Response to the critique of ``Cotton Gravity''}}
\author{R.A. Sussman$^{1}$, C.A. Mantica $^{2}$, L.G. Molinari $^{2}$ and
S. Najera $^{1}$}
\date{$^{1}$Institute of Nuclear Sciences, National Autonomous University
of Mexico, UNAM\\
 $^{2}$Physics Department Aldo Pontremoli, Università degli Studi
di Milano \\
 and I.N.F.N. sezione di Milano\\[2ex]
}
\maketitle
\begin{abstract}
Clement and Noiucer  submitted a note  {\tt arXiv:2401.16008 [gr-qc]} 
replying to our criticism {\tt arXiv:2401.10479 [gr-qc]} of their previous submission. 
We reply to the contents of this note and remark that these authors have not 
not addressed our arguments.  This will be our last response to them. Readers are advised to look at all material and judge by themselves
\end{abstract}

\section{Introduction}

Clement and Noiucer are correct in one issue mentioned in their note \cite{ClementNote}: we misunderstood their claim 
that the shortcomings of Cotton gravity become more severe as the theory is applied to spacetimes
with higher (not lower) symmetries. We apologize for this mistake. However, every single other argument we expressed in our criticism in \cite{response}  of their original submission \cite{Clement} remains
valid and has not been addressed nor disproved by Clement and Noiucer. 

The core argument posed by Clement and Nouicer to wish ``farewell'' to Cotton Gravity
is the underdetermination of the field equations in the Cotton formulation 
on static spherical symmetry and Bianchi I models, adding in \cite{ClementNote} the case of FLRW models. Clement and
Nouicer still claim that 
\begin{quote}
``{\it Actually, we did mention the Codazzi approach to Cotton gravity in our paper, 
stating that this approach being strictly equivalent to
the original Harada formulation of Cotton gravity, the under-determination
of the equations of Cotton gravity for highly symmetric configurations carries 
over to the under-determination of the Codazzi parametrization for the
same configurations.}''
\end{quote} 
This claim is incorrect. They did mention the Codazzi formulation in \cite{Clement}, but as we showed in detail in \cite{response}, they failed to understand its implications and utility.  We showed that, in spite of this ``strict equivalence'' (which 
we explain discuss ahead), the underdetermination and other 
shortcomings mentioned by Clement and Nouicer (in \cite{Clement} and in the present note) can be avoided by using
the Codazzi formulation in dealing with both of their test examples (static spherical 
symmetry and Bianchi I) and in finding more general non-trivial and self-consistent solutions not affected by this underdetermination  \cite{Mantica1,Mantica2,Sussman1,Sussman2}. 

\section{How far the ``strict equivalence'' argument goes?}

The ``strict equivalence'' of the Cotton and Codazzi formulation is a crucial argument 
of Clement and Nouicer, justifying their rejection of our argument that the Codazzi formulation 
makes a difference and avoids shortcomings they have highlighted in the Cotton formulation. 
However, this equivalence  of both formulations only occurs 
at top level of the field equations. It is easy to show the differences between 
the formulations when actually working with these equations. Readers are invited to compare:
\begin{itemize}
\item {\bf Cotton formulation}. It is the original formulation derived by Harada \cite{Harada1}, leads directly to 
\begin{equation} C_{abc}-8\pi M_{abc}=0,\label{Cotton}\end{equation}
where  $C_{abc}$  is the third order Cotton tensor and $M_{abc}$ is the 
generalized angular momentum. 
Both tensors must be computed at the onset for a given $g_{ab}$ and $T_{ab}$. 
Both $C_{abc}$ and $M_{abc}$ satisfy (as Clement and Nouicer mention) the 
constraints  $g^{ab}C_{abc}=g^{ab}M_{abc}=0$.  There are no further steps 
besides solving $C_{abc}=8\pi M_{abc}$, so already from the start 
there is no escape from the underdetermination they mention. Therefore,  
Clement and Nouicer announce their full dismissal of the theory.

\item {\bf Codazzi formulation}. It was derived by Mantica and Molinari \cite{Mantica1}. The tensors $C_{abc}$ and $M_{abc}$ are not computed from the onset, they only emerge at the end. 
The Codazzi formulation starts by proposing a nonzero second order tensor: 
\begin{equation}{\cal C}_{ab}=G_{ab}-8\pi T_{ab}-\frac13\left(G-8\pi T\right)\,g_{ab}\ne 0,
\label{Cod1}
\end{equation}
Notice that at this level there is no underdetermination and none of the problems with 
$C_{abc}$ and $M_{abc}$ appear, but there is a close connection 
to a correspondence with General Relativity (the case ${\cal C}_{ab}=0$).
Having found  ${\cal C}_{ab}\ne 0$ the next step is to demand that it is a Codazzi
tensor:
\begin{equation}\nabla_b{\cal C}_{ac}-\nabla_c{\cal C}_{ab}=0,\label{Cod2}
\end{equation}
Here is the ``strict equivalence'', since \eqref{Cod2} is identical to \eqref{Cotton}, but 
in the Codazzi formulation this equivalence appears at the end after an important
previous step involving a second order tensor ${\cal C}_{ab}\ne 0$, a step that never occurs in the Cotton formulation.  

 \end{itemize}
 
This comparison shows a striking contrast between working directly from the start with \eqref{Cotton}
(as Clement and Nouicer do) and working with the Codazzi formulation in \eqref{Cod1} followed by \eqref{Cod2}. Notice
that  with the Codazzi formulation there is no need to compute the constraints $g^{ab}C_{abc}=g^{ab}M_{abc}=0$,
since the end product is $C_{abc}=M_{abc} =0$ (when testing that ${\cal C}_{ab}\ne 0$ is a Codazzi tensor in \eqref{Cod2}). 

The approach using the Codazzi formulation works in practice, as we have proved by finding self-consistent non-trivial solutions in this formulation  \cite{Mantica1,Mantica2,Sussman1,Sussman2}, solutions that would be extremely difficult (or impossible) to obtain with the Cotton formulation.  Clement and Nouicer claim in \cite{ClementNote} that the theory (irrespective of its formulation) leaves the scale factor of FLRW models undetermined. This claim is disproved in \cite{Sussman1} and in \cite{Mantica2}, which used the Codazzi formulation. In fact, \cite{Mantica2} was not cited in \cite{Clement} and \cite{ClementNote}, but this work shows that only by using the Codazzi formulation it is possible  to reproduce the Friedman equations of many extended theories. In particular, in proposition 11 of \cite{Mantica2} it is proven the equivalence with Mimetic gravity, which is not regarded as an unphysical theory. 

Clement and Nouicer did mention in \cite{Clement} the Codazzi formulation and referred cursorily to solutions we obtained, but we find it regrettable that they simply glossed over our arguments and our articles on the Codazzi formulation without giving this work a minimal proper examination. Unfortunately, it seems that Clement and Nouicer are more concerned in dismissing the theory (``Farewell Cotton Theory'')  than in understanding it.


\begin{thebibliography}{10} 

\bibitem{ClementNote} G. Cl\'ement and K. Nouicer, \textit{Fairwell to Cotton gravity},  {\tt arXiv:2401.16008v1 [gr-qc]} 29 Jan 2024  

\bibitem{response} R.A. Sussman, C.A. Mantica, L.G. Molinari and S. Najera, {\it Response
to a critique of 'Cotton gravity'}, {\tt arXiv:2401.10479 [gr-qc]}

\bibitem{Clement} G. Cl\'ement and K. Nouicer, \textit{Cotton gravity
is not predictive}, {\tt arXiv:2312.17662v2 [gr-qc]} 01 Jan 2024

\bibitem{Mantica1} C.~A.~Mantica and L.~G.~Molinari, \textit{Codazzi
tensors and their space-times and Cotton gravity}, Gen. Relativ. Gravit.
\textbf{55} (4) 62, 2023.

\bibitem{Mantica2} C.~A.~Mantica, and L.~G.~Molinari, \textit{Friedmann
equations in the Codazzi parametrization of Cotton and extended theories
of gravity and the Dark Sector}, {\tt arXiv:2312.02784v2 {[}gr-qc{]}} 18
Dec 2023. To appear in {\it Phys Rev D}.

\bibitem{Sussman1} Sussman, R.A. and Najera, S. \textit{Cotton gravity:
the cosmological constant as spatial curvature},\\
 {\tt arXiv:2311.06744 [gr-qc]} 04 Dec 2023

\bibitem{Sussman2} Sussman, R.A. and Najera, S. \textit{Exact solutions
of Cotton Gravity in its Codazzi formulation, 2024}.\\
 {\tt  arXiv:2312.02115v2 [gr-qc]} 29 Jan 2024
 
 \bibitem{Harada1} Harada, J.\textit{ Emergence of the cotton tensor
for describing gravity}. Phys. Rev. D, \textbf{103} (12):L121502,
2021.


\end{thebibliography}
\end{document}